# Spontaneous laser line sweeping in bi-directional ring thulium fiber laser

Hongbo Jiang[1], Zihao Zhao[1], Pengtao Yuan[1], Lei Jin[2],Sze Yun Set[2]* and Shinji Yamashita[2]

*[1]Department of Electrical Engineering Information System, University of Tokyo, Bunkyo, Tokyo 113-8656, Japan*
*[2]Research Center on Advanced Science and Technology, University of Tokyo, Komaba, Tokyo 153-8904, Japan*

E-mail: hbjiang@cntp.t.u-tokyo.ac.jp; set@cntp.t.u-tokyo.ac.jp;

We report a phenomenon of self-sweeping in a bi-directional ring thulium-doped fiber laser, for the first time. The laser is spontaneously sweeping in both directions at a rate up to 0.2 nm/s with 15 nm sweeping range in 1.95 μm wavelength region. The laser output is switchable between two different working modes: periodical spontaneous laser line sweeping with generation of microsecond pulses in time domain; or static central wavelength with amplitude modulated temporally.

Spontaneous laser line sweeping (SLLS) was first observed in ruby laser [1] and re-discovered in ytterbium fiber laser in 2011 [2,3] when the terminology of this phenomenon was decided as well. Soon afterwards SLLS effect attracted more attention and achieved via various active medium include ytterbium, [4-7] erbium, [8,9] bismuth, [10] holmium,[11] and thulium. [12,13] This effect in fiber laser could be described as: The laser central wavelength keep drifting from the original lasing wavelength towards to the boundary of the sweeping range, and then bounce back after reaches the boundary. The physical mechanism of spontaneous sweeping of laser frequency was attributed to the formation of transient gratings along the active fiber which has been experimentally illustrated by P. Navratil [6] and I. Lobach [14] and explained by sophisticated simulation model as well. [7, 15 16] However, reports on SLLS and theoretical analysis were based on linear laser configuration, the SLLS effect in bi-directional ring fiber laser has not been reported yet. Despite of previous work by P. Peterka in 2011 with the customized wavelength selective element and optical isolator in a unidirectional Yb-doped ring fiber laser. [4] Bi-directional laser with counter propagating in clockwise (CW) and counter-clockwise (CCW) directions has been investigated for decades. [17,18] Different types of Bi-directional oscillation and laser frequency features have been studied in different types of such as gas ring lasers,[19 20] semiconductor ring lasers [21] and fiber ring lasers [22] in order to turn a bi-directional ring laser into real practical field like rotation sensing. Besides, the SLLS effect enables laser itself as a concise wavelength tunable source without external control therefore applicable in component interrogation and sensing system. A phase shift FBG characterization [23] and Brillouin optical spectrum analyzer [24] based on SLLS were reported and the demonstration of SLLS fiber laser water absorption sensor working in 2 μm was reported recently [13, 25]. Therefore it is worth to explore the impact of SLLS effect in bi-directional ring fiber laser.

In this paper we reported SLLS effect experimentally in a bi-directional ring Tm-doped fiber laser. The main feature of the laser is that dual working modes could be achieved by changing state of polarization in the cavity. The SLLS accompanied with pulses train generation in time domain was detected. In addition, the second working mode with static lasing wavelength and amplitude modulated temporally was observed as well. The observed phenomenon and experiment results will make original contributions to studies on longitudinal mode tuning, self-pulsing and bi-directional mode coupling in bi-directional ring fiber laser.

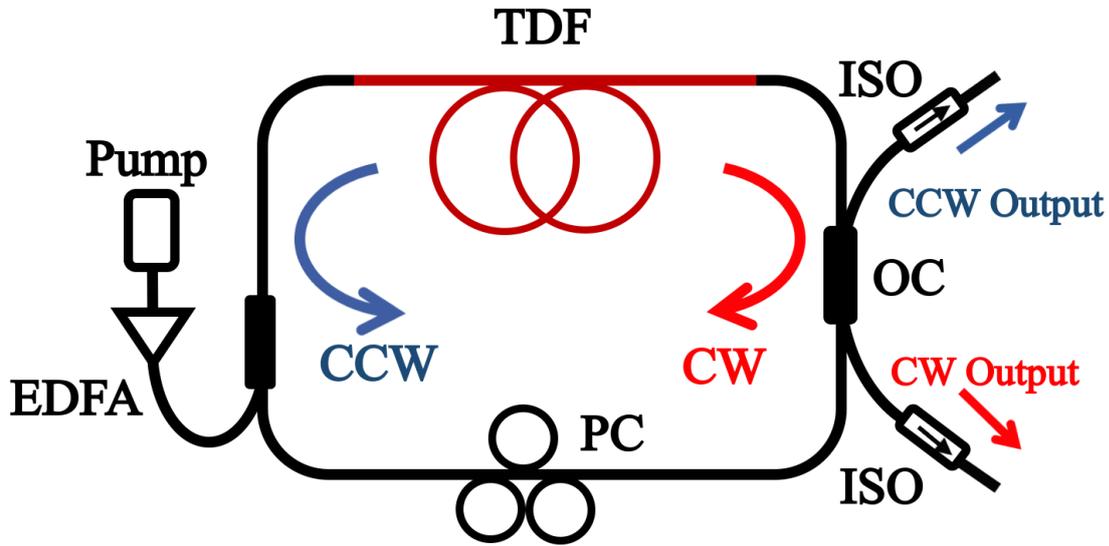

**Fig. 1.** Laser Setup. WDM: wavelength-division multiplexing; TDF: Thulium-doped fiber; OC: Output coupler; ISO: Isolator; PC: Polarization controller; CCW: Counter-clock wise; CW: Clock wise;

The experiment setup is shown in Fig. 1. The output from an EDFA working at 1570 nm was used as pump source for our Tm-doped fiber laser. 2.5 meters long thulium doped fiber was utilized as gain medium and connected with the cavity through 1570/2000 nm wavelength division multiplexing fused coupler. A four port optical fused coupler with 20% extracting ratio was implemented for bi-directional propagation. Optical isolators were used for blocking the unwanted reflection while the resonator was isolator-free. Polarization control was achieved by fiber coils type polarization controller inside the cavity. The cavity length was about 14 meters. Once pumping power reached the threshold about 120 mW and the polarization state was adjusted, optical spectrum of both direction started sweeping simultaneously.

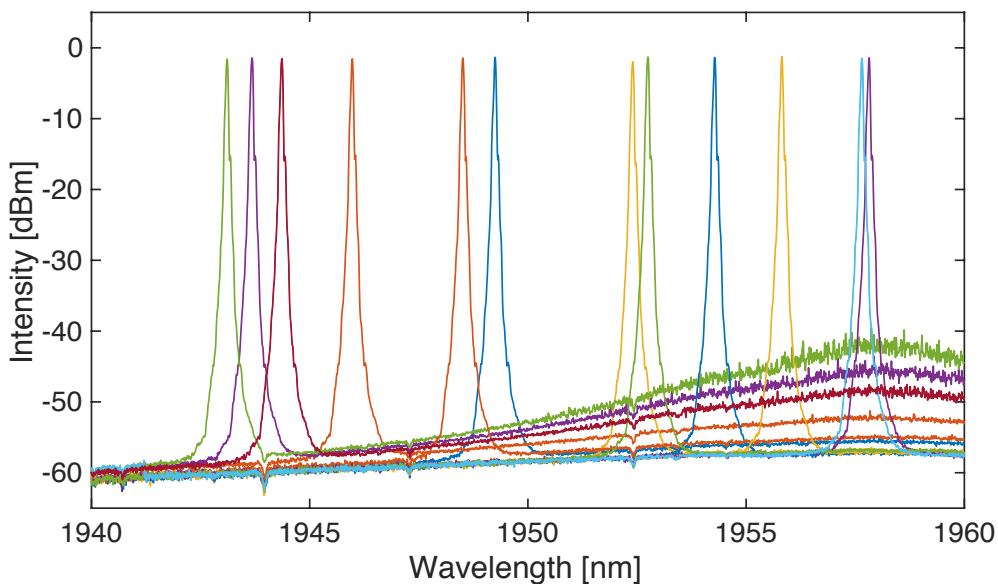

**Fig. 2.** Typical optical spectrum corresponding to the spontaneous laser line sweeping in single period

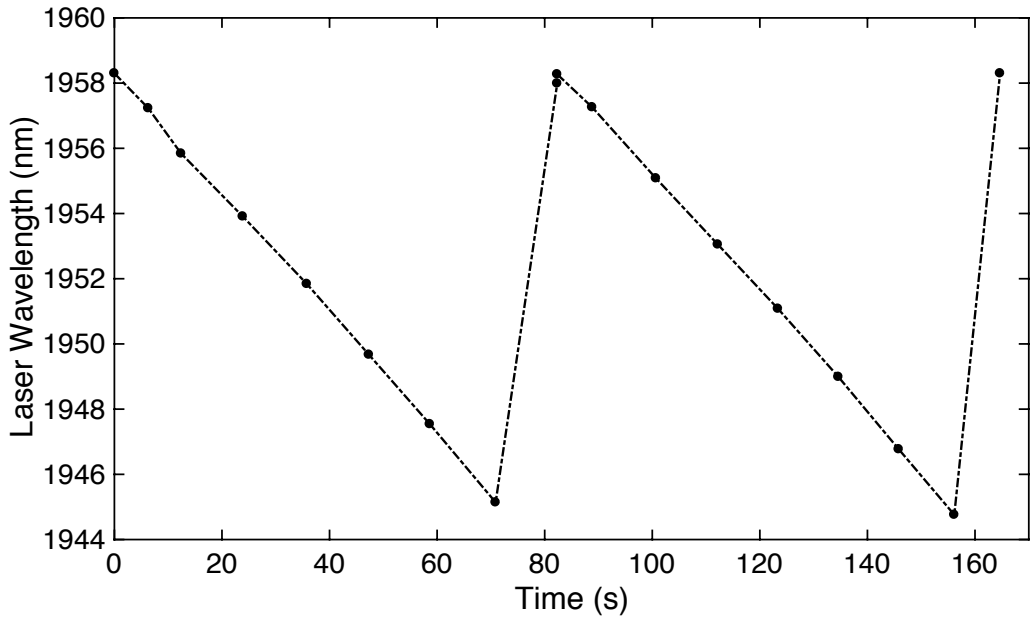

**Fig. 3.** Temporal dynamics of the output wavelength

The laser output spectra was detected by Optical Spectrum Analyzer (Yokogawa AQ6375). It is depicted in Fig. 2 that the sweeping range was about 15 nm. The wavelength was sweeping toward shorter wavelength with regular intensity until bounced back from most left side of sweep range periodically. It can be seen from Fig. 3 that the central wavelength of laser output was oscillating between 1960 nm and 1945 nm exhibited a rate of 0.2 nm/s with good long-term stability. We also noticed that pump level could affect sweeping range and sweep rate. When pump power increased, the range shrunk while the rate slightly increased. The latter result is in accordance with observations in [2, 12] but the range relationship is inconsistent with, [12] further investigation is needed in the future. Temporal behavior of outputs in both directions were measured by InGaAs photodetector (ET-5000F) and displayed with a digital oscilloscope. The results of outputs in time domain are shown in Fig. 4. Random microsecond pulses with frequency about 90~100 kHz were formed in both directions when the pump level was set to 220 mW. The pulse train was modulated by beating frequency between neighbor longitudinal modes, 14.5 MHz oscillation frequency is depicted in Fig. 4b, which corresponding to a 14.2 m-long laser cavity. The results of RF signal are presented in Fig. 5. The peak at 14.5 MHz coincided with the separation between adjacent longitudinal modes and its higher harmonics signal were observed in both directions. Expect for longitudinal mode beating (LMB) signal, the polarization beating signal (PMB) had regular distance with its neighbor LMB signal were detected as well. The PMB signal were varied with the state of polarization, and even disappeared when polarization controller was adjusted to specific position corresponds to the case that the SOP of counter-propagating waves were orthogonal. [26] In fact, when PMB signals disappeared, the laser wavelength was no longer sweeping. In the other words, the laser working mode could be transferred from sweeping to static.

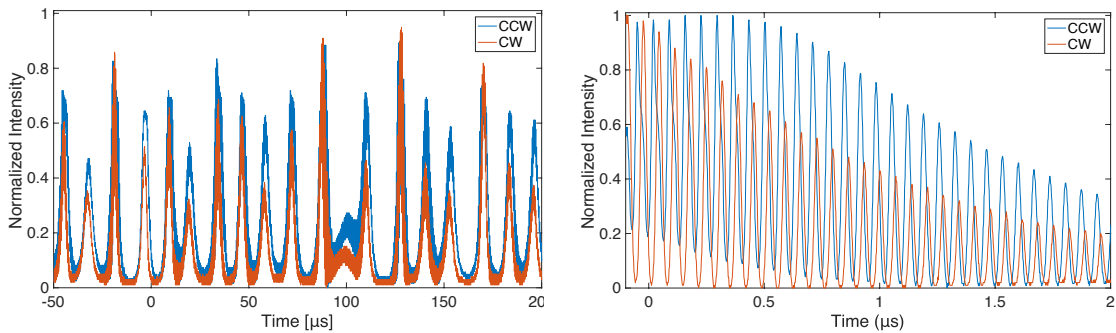

**Fig. 4.** Pulses train (a) and intermode beating inside the pulse (b) in both directions when laser wavelength is sweeping

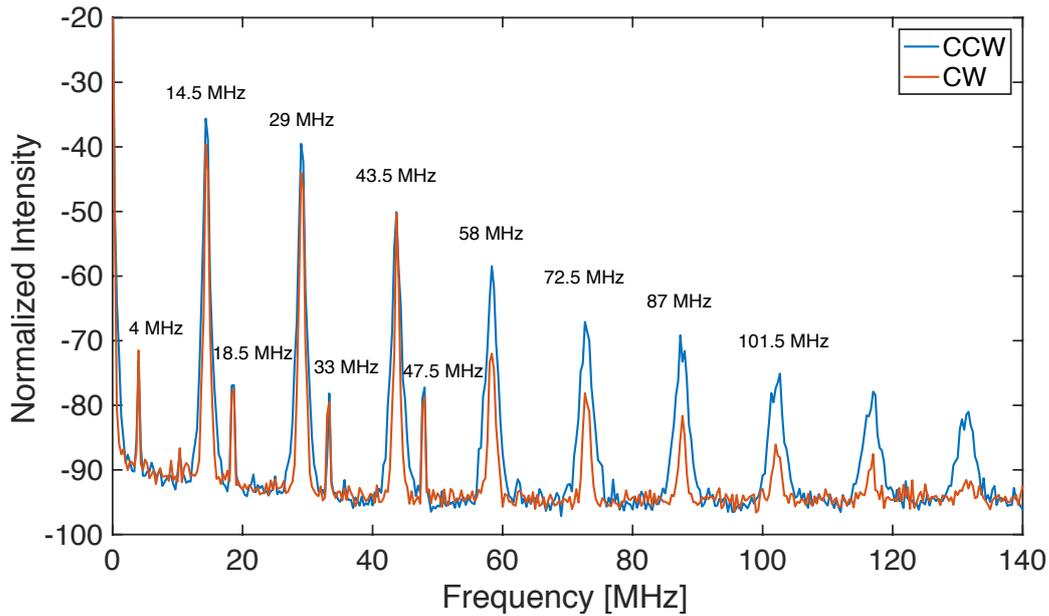

**Fig. 5.** Longitudinal mode beating signals and Polarization beating signals

The mechanism in bi-directional ring fiber laser about SLLS effect can be analyzed briefly. It has been well known that mode coupling between CW and CCW due to scattering from the grating formed in the gain medium [7,15,21] and back-reflections from imperfect facets [4, 27] will determine the energy exchange and oscillation feature between two directions. [27, 28] These mode coupling lead to variation of population inversion, then leads to the spatial variation of gain and refractive index along active medium [7, 15] owning to Kramers-Kronig relation [16, 29] therefore the transient gratings formed. In addition, the gratings written by longitudinal mode m is suppressed so lasing frequency hops to another mode constantly.

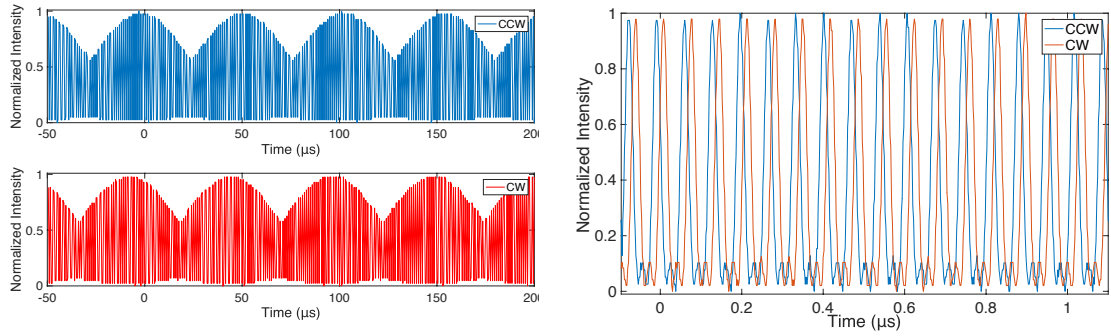

**Fig. 6.** Amplitude modulated (a)and intermode beating (b) in both directions

The self-pulsing temporal features in Fig. 4 was attributed to the relaxation oscillation since each longitudinal mode emitted and existed for a limited time [30]. The discussions on sweeping direction, integer between pervious mode and current mode and reflectivity of these gratings will be extended in future works.

In contrast to previous reports on SLLS effect which only observed one working mode [1-12, 14] or the transition between two modes is unstable.[13] Our results record dual working modes with good long-term stability individually. Switching between two working modes is achieved by adjusting polarization controller. The static mode is different from sweeping mode temporally more than obvious spectral distinction. The oscillatory output under static working mode in both directions is depicted in Fig. 6. The amplitude-modulated results is consistent with theoretical work by R. Neelen [31] in 1991 which assuming mode amplitudes modulation was attributed to large enough linear backscattering coefficient.

In conclusion, we report a spontaneous laser wavelength sweeping in bi-directional ring Tm fiber laser. The laser is operated laser wavelength periodical sweeping mode with sweep range up to 18 nm in both directions without any external control. The second working mode with static laser wavelength is achieved by only changing the state of polarization. The results are helpful for further investigation of mode coupling behavior in CCW-CW ring fiber laser, which will arise attention for turning a bi-directional ring fiber laser into practical fields.

# References

1) V V Antsiferov and V S Pivtsov and V D Ugozhaev and Kim G Folin, 1973 Sov. J. Quantum Electron. 3 211.

2) Ivan A. Lobach, Sergey I. Kablukov, Evgeniy V. Podivilov, and Sergey A. Babin, Opt. Express 19, 17632-17640 (2011).

3) A V Kir'yanov and N N Il'ichev 2011 Laser Phys. Lett. 8 305

4) P Peterka, P Navrátil, J Maria, B Dussardier, R Slavík, P Honzátko and V Kubeček, 2012 Laser Phys. Lett. 9 445

5) P Peterka, J Maria, B Dussardier, R Slavík, P Honzátko and V Kubeček, 2009 Laser Phys. Lett. 6 732

6) P Navratil, P Peterka, P Honzatko and V Kubecek, 2017 Laser Phys. Lett. 14 035102

7) I A Lobach, S I Kablukov and E V Podivilov and S A Babin, 2014 Laser Phys. Lett. 11 045103

8) P. Navratil, P. Peterka, P. Vojtisek, I. Kasik, J. Aubrecht, P. Honzatko, V. Kubecek, Opto-Electronics Review 26 (2018) 29-34

9) P. Navratil, P. Vojtíšek, P. Peterka, P. Honzátko, V. Kubeček, Proc.SPIE, 8697, 86971M (18 December 2012)

10) Ivan A. Lobach, Sergey I. Kablukov, Mikhail A. Melkumov, Vladimir F. Khopin, Sergey A. Babin, and Evgeny M. Dianov, Opt. Express 23, 24833-24842 (2015)

11) Holium fiber laser near 2100nm Jan Aubrecht, Pavel Peterka, Pavel Koška, Ondřej Podrazký, Filip Todorov, Pavel Honzátko, and Ivan Kašík, Opt. Express 25, 4120-4125 (2017)

12) Xiong Wang, Pu Zhou, Xiaolin Wang, Hu Xiao, and Lei Si, Opt. Express21, 16290-16295 (2013)

13) A. E. Budarnykh, A. D. Vladimirskaya, I. A. Lobach, and S. I. Kablukov, Opt. Lett. 43, 5307-5310 (2018)

14) I. A. Lobach, R. V. Drobyshev, A. A. Fotiadi, and S. I. Kablukov, in *Frontiers in Optics 2016*, OSA Technical Digest (online) (Optical Society of America, 2016), paper FTu2I.6.

15) P. Peterka, P. Honzátko, P. Koška, F. Todorov, J. Aubrecht, O. Podrazký, and I. Kašík, Opt. Express 22, 30024-30031 (2014)

16) P. Peterka, P. Koška and J. Čtyroký, IEEE Journal of Selected Topics in Quantum Electronics, vol. 24, no. 3, pp. 1-8, May-June 2018, Art no. 902608.

17) H. Haus, H. Statz and I. Smith, IEEE Journal of Quantum Electronics, vol. 21, no. 1, pp.

78-85, January 1985.

18) T. H. Chyba, Phys Rev A Gen Phys. 1989 Dec 1;40(11):6327-6338.

19) W. R. Christian and L. Mandel, J. Opt. Soc. Am. B 5, 1406-1411 (1988)

20) Spreeuw, R. J. C. and Neelen, R. Centeno and van Druten, N. J. and Eliel, E. R. and Woerdman, J. P., PhysRevA.42.4315 (1990)

21) Neelen, R.Centeno & Van Exter, M.P. & Bouwmeester, D & Woerdman, J.P., Journal of Modern Optics. August 1992. 1623-1641.

22) Roman Kiyan, Seung Kwan Kim and Byoung Yoon Kim, IEEE Photonics Technology Letters, vol. 8, no. 12, pp. 1624-1626, Dec. 1996.

23) Ivan A. Lobach and Sergey I. Kablukov, J. Lightwave Technol. 31, 2982-2987 (2013)

24) A. Yu. Tkachenko, I. A. Lobach, and S. I. Kablukov, Opt. Express 25, 17600-17605 (2017)

25) A. E. Budarnykh, A. D. Vladimirskaya, I. A. Lobach, S. I. Kablukov, Proc. SPIE 10814, 108140R (5 November 2018);

26) S. K. Kim, H. K. Kim, and B. Y. Kim, Opt. Lett. 19, 1810-1812 (1994)

27) Wen Ji, Shufen Chen, Lei Fu, and Zhengfeng Zou, Opt. Lett.37, 3588-3590 (2012)

28) Fan-Jun Rao, Shu-fen Chen, Lei Fu, Optics Communications, Volume 284, Issue 5,2011

29) Henrik Tünnermann, Jörg Neumann, Dietmar Kracht, and Peter Weßels, Opt. Express 20, 13539-13550 (2012)

30) Simin Wang , Wei Lin, Weicheng Chen, Can Li, Changsheng Yang, Tian Qiao and Zhongmin Yang, Applied Physics Express 1882-0786-9-3-032701(2016)

31) R. Centeno Neelen, R. J. C. Spreeuw, E. R. Eliel, and J. P. Woerdman, "Frequency splitting of the longitudinal modes of a ring dye laser due to backscattering," J. Opt. Soc. Am. B 8, 959-969 (1991)